\pdfoutput=1
\RequirePackage{ifpdf}
\ifpdf % We are running pdfTeX in pdf mode
\documentclass[pdftex]{sigma}
\else
\documentclass{sigma}
\fi

\newcommand{\re}{\operatorname{Re}}
\newcommand{\im}{\operatorname{Im}}

\numberwithin{equation}{section}

\begin{document}
%\allowdisplaybreaks

\newcommand{\arXivNumber}{1804.03039}

\renewcommand{\PaperNumber}{092}

\FirstPageHeading

\ShortArticleName{Maximally Superintegrable Systems in a Magnetic Field with Higher Order Integrals}

\ArticleName{An Infinite Family of Maximally Superintegrable\\ Systems in a Magnetic Field with Higher Order\\ Integrals}

\Author{Antonella MARCHESIELLO~$^\dag$ and Libor {\v{S}}NOBL~$^\ddag$}

\AuthorNameForHeading{A.~Marchesiello and L.~{\v{S}}nobl}

\Address{$^\dag$~Czech Technical University in Prague, Faculty of Information Technology,\\
\hphantom{$^\dag$}~Department of Applied Mathematics, Th\'{a}kurova 9, 160 00 Prague 6, Czech Republic}
\EmailD{\href{mailto:marchant@fit.cvut.cz}{marchant@fit.cvut.cz}}

\Address{$^\ddag$~Czech Technical University in Prague, Faculty of Nuclear Sciences and Physical Engineering, \\
\hphantom{$^\ddag$}~Department of Physics, B\v rehov\'a 7, 115 19 Prague 1, Czech Republic}
\EmailD{\href{mailto:Libor.Snobl@fjfi.cvut.cz}{Libor.Snobl@fjfi.cvut.cz}}

\ArticleDates{Received April 10, 2018, in final form August 24, 2018; Published online August 31, 2018}

\Abstract{We construct an additional independent integral of motion for a class of three dimensional minimally superintegrable systems with constant magnetic field. This class was introduced in~[\textit{J.~Phys.~A:
 Math. Theor.} \textbf{50} (2017), 245202, 24~pages] and it is known to possess periodic closed orbits. In the present paper we demonstrate that it is maximally superintegrable. Depending on the values of the parameters of the system, the newly found integral can be of arbitrarily high polynomial order in momenta.}

\Keywords{integrability; superintegrability; higher order integrals; magnetic field}

\Classification{37J35; 78A25}

\section{Introduction}\label{sec: Introduction}

In our recent paper~\cite{MS} we found a class of minimally superintegrable systems in three spatial dimensions with constant magnetic field which possesses closed bounded periodic trajectories for a particular choice of parameters. Namely, a quantity $\kappa$ constructed out of them (cf.\ equation~\eqref{Sk} below) must be rational, $\kappa=\frac mn$. For three particular choices of $\kappa$ this system is known to be maximally superintegrable with integrals of at most second order in momenta~\cite{MS}. Thus a~natural question arises asking whether also for the remaining values of the parameters satisfying the rationality constraint a missing independent integral can be constructed.

In this paper we describe how the considered system can be reduced to the two dimensional anisotropic harmonic oscillator and how the known integrals of the anisotropic oscillator give rise to a new integral for the system with the magnetic field. Assuming we have $\kappa=\frac{m}{n}$ where~$m$ and~$n$ are incommensurable, the additional integral is of order $m+n-1$ in momenta. Its leading order terms involve angular momenta linearly.

For the sake of clearness, let us recall that in classical mechanics (with or without magnetic field) integrability means that there exist~$N$ integrals of motion~$X_j$ including the Hamiltonian that Poisson commute pairwise, are well defined functions on the phase space and are functionally independent. The system is superintegrable if it allows $k$ further independent integrals~$Y_a$ that Poisson commute with the Hamiltonian, but not necessarily with each other, nor with the integrals $X_j$. The integer $k$ satisfies $1\leq k \leq N-1$ where $k=1$ and $k=N-1$ correspond to minimal and maximal superintegrability, respectively. Similarly, in quantum mechanics we assume that the integrals are well-defined commuting self-adjoint operators, polynomial in the operators $\hat{x}_j$ and $\hat{p}_j$ representing the coordinates and the momenta, or more general objects, such as convergent series in these operators. Requirement of independence in this case means that no nontrivial fully symmetrized polynomial in the integrals of motion should vanish.

Maximally superintegrable systems are of special interest in classical physics because all finite trajectories in these systems are closed in configuration space and the motion is periodic. In quantum mechanics the energy levels are degenerate and it has been conjectured that maximally superintegrable systems are exactly solvable~\cite{TTW}.

Most of the recent research on superintegrability focused on ``natural'' Hamiltonians with scalar potentials. For early systematic study in three dimensions see~\cite{Evans1,Evans2,MaSmoVaWin,VerEv}. This paper belongs to a series of papers~\cite{MS,MSW,MSW2} studying superintegrability of three dimensional systems with magnetic fields. We refer the reader to the papers~\cite{MSW,MSW2} for discussion of a general motivation for our research and to~\cite{MS} for details concerning the introduction of the system considered in this paper. For a general discussion of integrability and superintegrability see a~recent review~\cite{MiPoWin}, for superintegrability in the presence of vector potentials, e.g., \cite{BeWin,DoGraRaWin,LaMayVi1,McICis,McSWin,Pucacco,PuRos,Zhalij}. The influence of constant magnetic field on the motion in a given potential was studied, e.g., in~\cite{TurEsc1,TurEsc2} where the two body Coulomb problem was studied. Despite the similarities the reader should notice the differences between the two approaches: in~\cite{TurEsc1} additional, so called particular, integrals conserved only on special ``superintegrable'' trajectories were constructed whereas we consider the appearance of additional global integrals when a particular relation among the parameters of the system holds.

The structure of the paper is as follows: In Section~\ref{sec: System} we describe our system, its integrals and its trajectories as presented in~\cite{MS}. In Section~\ref{sec: Reduction} we reduce its dynamics to the well-known case of two dimensional anisotropic harmonic oscillator (cf.~\cite{JauHi,Nakagawa,RTW}). In Section~\ref{sec: fifth integral} we use the known integrals of the anisotropic oscillator to construct previously unknown integrals for the three dimensional problem under investigation. In Section~\ref{sec: example} we present an explicit example. In the last section we conclude with a summary of our results and comment on the superintegrability of the quantum analogue of our system.

\section{The system}\label{sec: System}
We consider the Hamiltonian system on the phase space with coordinates $(\vec x,\vec p)$ where $\vec x=(x,y,z)$ and $\vec p=(p_1,p_2,p_3)$. We assume that its magnetic field and effective potential are given by
\begin{gather}\label{mageff}
\vec B(\vec x)=(-\Omega_1,\Omega_2,0),\qquad W(\vec x)=\frac{\Omega_1\Omega_2}{2S} (Sx-y)^2,
\end{gather}
where $\Omega_1$, $\Omega_2$, $S$ are real constants such that $S\neq 0$ and $\Omega_1^2+\Omega_2^2\neq 0$. Its Hamiltonian can be written as
\begin{gather}\label{classham}
H= \frac{1}{2} \big(\vec p^A\big)^2 + W(\vec x),
\end{gather}
where
\begin{gather*}
p_j^A=p_j+ A_j(\vec x)
\end{gather*}
are gauge covariant expressions for the momenta (for our choice of mass equal to 1 they coincide with the velocities). We notice that in the usual formulation of Hamiltonian dynamics which we shall use here for computational reasons, the equations of motion on the phase space and thus also the integrals of motion expressed in terms of the canonical variables $x_i$ and $p_j$ depend on the chosen gauge. However, the existence of the integrals as well as their expression in terms of the coordina\-tes~$x_i$ and the covariantized momen\-ta~$p_j^A$ does not depend on the choice of gauge. This is best seen from the fact that the (non-canonical) Poisson brackets among the coordina\-tes~$x_i$ and~$p_j^A$
\begin{gather*}
\{x_i,x_j\}_{\rm P.B.}=0,\qquad \big\{x_i,p_j^A\big\}_{\rm P.B.}= \delta_{ij}, \qquad \big\{p_i^A,p_j^A\big\}_{\rm P.B.}= -\epsilon_{ijk} B_k
\end{gather*}
as well as Poisson bracket of any functions expressed in terms of them are explicitly gauge invariant (see, e.g., \cite[Remark~5.1.10(6), p.~217]{Thirring}). Thus the notions of integrability and superintegrability are gauge-invariant.

The system \eqref{mageff} is already known to be minimally superintegrable \cite{MS}. It has three Cartesian type integrals
\begin{gather}
\nonumber X_0 = p_3^A+\Omega_2 x+\Omega_1 y, \\
X_1 = \big(p_1^A\big)^2-2 \Omega_2 x p_3^A-\Omega_2^2 x^2+\Omega_1\Omega_2 x(Sx-2 y), \nonumber\\ X_2 =\big( p_2^A\big)^2-2 \Omega_1 y p_3^A-\Omega_1^2y^2+\frac{\Omega_1\Omega_2}{S} y(y-2 Sx)\label{X1X2}
\end{gather}
on which the Hamiltonian is polynomially dependent, $ H=\frac{1}{2}\big( X_0^2+X_1+X_2 \big)$, and an additional first order integral
\begin{gather}\label{X3}
X_3=p_1^A+S p_2^A- (S\Omega_1+\Omega_2)z.
\end{gather}

The classical trajectories of the system~\eqref{mageff} are known, cf.~\cite{MS}
\begin{gather}
\nonumber x(t) = \frac{1}{\omega_1^2}\big( \big( \omega_1^2 x_0 - \Omega_2 p_{30}\big)\cos ( \omega_1 t ) + \omega_1 p_{10} \sin (\omega_1 t ) + \Omega_2 p_{30} \big),\\
\nonumber y(t) = \frac{1}{\omega_2^2} \big(\big( \omega_2^2 y_0- \Omega_1 p_{30} \big)\cos (\omega_2 t)+ \omega_2 p_{20} \sin (\omega_2 t )+ \Omega_1 p_{30} \big), \\
\nonumber z(t) = \frac{1}{\Omega_1 S+\Omega_2} \bigg( p_{10}( \cos(\omega_1t) -1)+ S p_{20} ( \cos(\omega_2 t)-1) \\
\hphantom{z(t) =}{} +\frac{\Omega_2 p_{30} -\omega_1^2 x_0}{\omega_1} \sin(\omega_1t) +\frac{ \Omega_1 p_{30} - \omega_2^2 y_0 }{ \omega_2} \sin(\omega_2 t) \bigg)+z_0,\label{classsol}
\end{gather}
where we introduced the constants
\begin{gather}\label{omegas}
\omega_1=\sqrt{\Omega_2(\Omega_1 S+\Omega_2)}, \qquad \omega_2=\sqrt{\frac{\Omega_1}{S} ( \Omega_1S+\Omega_2 )}=\sqrt{\frac{\Omega_1}{S \Omega_2}}\omega_1
\end{gather}
in order to shorten the terms in~\eqref{classsol}.

In \cite{MS} it was also proven that in the special cases
\begin{gather*}
S=\frac{\Omega_1}{\Omega_2},\qquad S=4\frac{\Omega_1}{\Omega_2} \qquad \text{and}\qquad S=\frac{\Omega_1}{4\Omega_2}
\end{gather*}
the system~\eqref{mageff} is maximally superintegrable, with the additional integral of order $1$ and $2$, respectively. In the following we prove that the system~\eqref{mageff} is maximally superintegrable whenever the trajectories~\eqref{classsol} are periodic (or, equivalently, closed), i.e., for
\begin{gather} \label{Sk}
S=\frac{\Omega_1}{\Omega_2}\kappa^2,\qquad \text{where} \quad \kappa=\frac mn,\quad m,n\in\mathbb{N} \ \text{are} \ \text{incommensurable},
\end{gather}
with the fifth integral of order $ m + n -1$ in the momenta $p_1$, $p_2$, $p_3$. We notice that systems with the parameters $\Omega_1$, $\Omega_2$, $\kappa$ and $\Omega_2$, $\Omega_1$, $\frac{1}{\kappa}$ are equivalent, differ just by a choice of Cartesian coordinates, cf.~\cite{MS}.

We present the system as expressed in~\eqref{mageff} since this is the form in which its mathematical structure is most easy to analyze. However, this is not the best point of view for its physical interpretation. Through a rotation the system can be brought to a form where either the magnetic field $\vec B$ or the harmonic potential $W$ is aligned with one of the coordinate axis. In~\cite{MS} we demonstrated its form when the magnetic field is aligned with the $x$-axis; however it may be more illuminating to rotate it so that the potential $W$ acts along a coordinate axis. Thus we rotate our coordinate system about the $z$-axis by an angle $\alpha$ such that $\tan \alpha=\frac{1}{S}= \frac{\Omega_2}{\kappa^2 \Omega_1}$. The system takes the form~\eqref{classham} with the harmonic oscillator potential acting along one coordinate axis,
\begin{gather}\label{tildeW}
\tilde W(\tilde x,\tilde y,\tilde z)= \frac{1}{2} \left( \kappa^2 \Omega_1^2+\frac{\Omega_2^2}{\kappa^2} \right) \tilde x^2\equiv \frac{1}{2} (\hat \omega)^2 \tilde x^2
\end{gather}
in the constant magnetic field
\begin{gather*}
\vec{\hat{B}}=\frac{1}{\sqrt{\Omega_2^2+\kappa^4 \Omega_1^2}}\big( {-}\kappa^2 \Omega_1^2-\Omega_2^2, \Omega_1 \Omega_2 \big(\kappa^2-1\big) ,0\big)\equiv \big(\hat B \cos \beta,\hat B \sin \beta,0\big).
\end{gather*}
Vice versa, we may, at least on some open neighborhood, express $\Omega_1$, $\Omega_2$ and $\kappa$ in terms of the three new parameters $\hat \omega$, $\hat B$ and~$\beta$ by solving algebraic equations. Such three dimensional unidirectional harmonic oscillator inserted in a constant magnetic field is minimally superintegrable with two first order integrals~$\tilde{p}_2^A$ and $\tilde{p}_3^A-\hat B y\cos\beta +\hat B x\sin \beta$ and one independent second order integral. We see that the integral~\eqref{X3} reflects the invariance of the system under translation along the $\tilde y $ direction. The solutions of the equations of motion exhibit oscillatory behaviour with two independent frequencies
\begin{gather*}
\omega_1 = \frac{1}{\sqrt{2}}\sqrt{\hat B^2+\hat{\omega}^2+\sqrt{\big(\hat B^2+\hat{\omega}^2\big)^2-4 \hat B^2 \hat{\omega}^2 \cos(\beta)^2}},\\
\omega_2 = \frac{1}{\sqrt{2}}\sqrt{\hat B^2+\hat{\omega}^2-\sqrt{\big(\hat B^2+\hat{\omega}^2\big)^2-4 \hat B^2 \hat{\omega}^2 \cos(\beta)^2 }}\end{gather*}
(they coincide with the frequencies $\omega_{1,2}$ in~\eqref{omegas}). When these frequencies become resonant,
\begin{gather*}
\frac{\omega_1}{\omega_2}=\kappa=\frac{m}{n},
\end{gather*} a new higher order integral to be constructed below appears.

\section{Reduction to the anisotropic oscillator}\label{sec: Reduction}
Oscillator potentials and constant magnetic fields share some similarities, as noticed, e.g., in~\cite{DehHus}. On the other hand, the first order integrals $X_0$ and $X_3$ in \eqref{X1X2} and \eqref{X3} show that the system can be regarded as an oscillator in the direction of $\tilde x$, cf.~\eqref{tildeW} plus a constant magnetic field. In the following we see how, by using the known first order integrals, the system \eqref{mageff} can indeed be reduced to a two dimensional anisotropic oscillator for the parameter $S$ satisfying~\eqref{Sk}. With gauge chosen as
\begin{gather}\label{CaseA2vecpot}
\vec A=(0,0,-\Omega_2 x-\Omega_1 y),
\end{gather}
the Hamilton equations read
\begin{gather*}
\dot x=p_1,\qquad \dot y=p_2,\qquad \dot z=-\Omega_1 y-\Omega_2 x +p_3,\qquad \dot p_3=0,\nonumber\\
 \dot p_1=-\big(\Omega_2^2 +\kappa^2 \Omega_1^2\big) x+ \Omega_2 p_3,\qquad \dot p_2=-\left(\Omega_1^2 + \frac{ \Omega_2^2 }{\kappa^2}\right)y+ \Omega_1 p_3.%\label{HJ1}
\end{gather*}
By substituting $p_3\equiv p_{30}$ and by the shift
\begin{gather}\label{shift}
 x= X+\frac{\Omega_2 p_{30}}{\Omega_2^2+ \Omega_1^2 \kappa^2},\qquad y=Y+\frac{\Omega_1 p_{30} \kappa^2}{\Omega_2^2+ \Omega_1^2 \kappa^2},
\end{gather}
the equations simplify to
\begin{gather}\label{H2J2a}
\dot P_1=-\left(\Omega_2^2+ \kappa^2 \Omega_1^2\right) X,\qquad \dot P_2=-\left(\Omega_1^2 + \frac{ \Omega_2^2 }{\kappa^2}\right)Y,\qquad \dot X=P_1, \qquad \dot Y=P_2,\\
 \dot z = -\Omega_1 Y-\Omega_2 X +p_{30},\label{H2J2b}
\end{gather}
where $P_1$, $P_2$ are the momenta conjugated to the new space coordinates $X$, $Y$ (once evaluated they are equal to the original~$p_1$,~$p_2$). By solving the first two equations in~\eqref{H2J2a} with respect to~$X$ and~$Y$ and substituting into~\eqref{H2J2b} we find
\begin{gather*} \Omega_2\dot P_1 + \Omega_1 \kappa^2 \dot P_2 - \big(\Omega_2^2 + \Omega_1^2 \kappa^2\big) \dot z=0
\end{gather*}
corresponding to the integral~\eqref{X3}. The dynamics are thus reduced to the dynamics of an anisotropic oscillator, whose frequency ratio is $\kappa$ and canonical coordinates are $(X,Y,P_1,P_2)$. It is known~\cite{RTW} that if $\kappa$ satisfies~\eqref{Sk}, such an oscillator is superintegrable. Let us henceforth restrict to this case and set
\begin{gather*}
\omega^2= \frac{\Omega_1^2}{n^2}+ \frac{\Omega_2^2}{m^2}.
\end{gather*}
The Hamiltonian of the two degree of freedom (from now on abbreviated to d.o.f.) oscilla\-tor~\eqref{H2J2a} is obtained by substituting~\eqref{shift} into the Hamiltonian~\eqref{classham} and reads
\begin{gather} \label{H2dof}
H_2=\frac12\big(P_1^2+P_2^2\big)+ \frac12 \omega^2\big(m^2 X^2+n^2 Y^2\big).
\end{gather}
By introducing complex coordinates
$z_1= i P_1+m \omega X$, $z_2= i P_2+n \omega Y$, the ring of the invariants of the oscillator \eqref{H2dof} is generated by \cite{JauHi,RTW}
\begin{gather}
I_1 = z_1\bar z_1,\qquad I_2=z_2\bar z_2, \qquad I_3= \operatorname{Re} \big(z_1^n \bar z_2^m\big),\qquad I_4= \operatorname{Im} \big(z_1^n \bar z_2^m\big).\label{I34}
\end{gather}
The invariants~\eqref{I34} are clearly not independent; they satisfy the relation
\begin{gather*}%\label{Invariant Rel}
I_3^2+ I_4^2= I_1^n I_2 ^m.
\end{gather*}
Equivalently, the integrals $I_3$, $I_4$ can be expressed in terms of Chebyshev polynomials as~\cite{EvaVe}
\begin{gather}
\nonumber I_3 = |z_1|^{n-1} |z_2|^{m-1}\left( |z_1| |z_2| T_n\left(\frac{\re{z_1}}{|z_1|}\right)T_m\left(\frac{\re{z_2}}{|z_2|}\right)\right.\nonumber\\
\left. \hphantom{I_3 =}{} + \im{z_1}\im{z_2} U_{n-1}\left(\frac{\re {z_1}}{|z_1|}\right) U_{m-1}\left(\frac{\re z_2}{|z_2|}\right) \right),\nonumber\\
I_4 = |z_1|^{n-1} |z_2|^{m-1}\left(|z_2|\im{z_1} U_{n-1}\left(\frac{\re z_1}{|z_1|}\right) T_m\left(\frac{\re{z_2}}{|z_2|}\right)-\right.\nonumber\\
\left.\hphantom{I_4 =}{} - |z_1|\im z_2 T_n\left(\frac{\re z_1}{|z_1|}\right) U_{m-1}\left(\frac{\re{z_2}}{|z_2|}\right) \right),\label{I3I4Chebyshev}
\end{gather}
where $T_n$, $U_n$ denote the Chebyshev polynomial of degree $n$ of the first and second type, respectively.

As we will show in the next section, the integrals $I_1$ and $I_2$ correspond to the Cartesian type integrals $X_1$ and $X_2$ of the original system while $I_3$ (or $I_4$) gives a new independent integral for the system~\eqref{mageff}. We find it interesting to note that, under the assumption \eqref{Sk}, the system reduces to the Landau problem (i.e., a particle moving in a constant magnetic field without any potential force) in the limit $\Omega_1\rightarrow0$ if~$S$ is kept constant, i.e., $\kappa\rightarrow+\infty$, $\Omega_1 \approx \frac{1}{\kappa^2}$. This reflects the fact that one of the frequencies in the oscillator~\eqref{H2dof} goes to zero, and the polynomial integrals~\eqref{I3I4Chebyshev} are becoming of increasing order until they are lost in the limit.

\section{The fifth integral} \label{sec: fifth integral}
Let us first invert the shift \eqref{shift}. The integrals $I_1$ and $I_2$ give, after neglecting terms proportional to $p_{30}^2$,
\begin{gather}
\tilde I_1 = p_1^2 -2\Omega_2 x p_{30} +\frac{\Omega_1^2 m^2 +\Omega_2^2 n^2}{n^2}x^2,\label{I1shiftinv}\\
\tilde I_2 = p_2^2 -2\Omega_1 y p_{30} +\frac{\Omega_1^2 m^2 +\Omega_2^2 n^2}{m^2} y^2.\label{I2shiftinv}
\end{gather}
By substituting back $p_{3}=p_{30}$ into \eqref{I1shiftinv}--\eqref{I2shiftinv} we see they correspond to the Cartesian type integrals $X_1$ and $X_2$ of \eqref{mageff}.

Similarly, we can find the expressions of the integrals $I_3$ and $I_4$ in the original phase space coordinates.
We find it convenient to work with the following series expressions for Chebyshev polynomials
\begin{gather*}
T_n(a)=\sum_{k=0}^{[\frac{n}{2}]}\binom{n}{2k} a^{n-2k}\big(a^2-1\big)^k,\qquad
U_n(a)=\sum_{k=0}^{[\frac{n}{2}]}\binom{n+1}{2k+1} a^{n-2k}\big(a^2-1\big)^k,
\end{gather*}
so that we can explicitly write the two integrals as polynomials in the momenta. Namely, after inverting the shift \eqref{shift} and substituting $p_{30}=p_3$ we have, in gauge covariant form,
\begin{gather}
X_4 = \sum_{k=0}^{[\frac{n-1}{2}]}(-1)^k\binom{n}{2k+1} \big(m \omega \tilde X^A\big)^{n-2k-1}\big(p_1^A\big)^{2k+1}\nonumber\\
\hphantom{X_4 =}{} \times \sum_{k=0}^{[\frac{m-1}{2}]}(-1)^k\binom{m}{2k+1} \big(n \omega \tilde Y^A\big)^{m-2k-1}\big(p_2^A\big)^{2k+1}\nonumber\\
\hphantom{X_4 =}{} + \sum_{k=0}^{[\frac{n}{2}]}(-1)^k\binom{n}{2k} \big(m \omega \tilde X^A\big)^{n-2k}\big(p_1^A\big)^{2k}\sum_{k=0}^{[\frac{m}{2}]}(-1)^k\binom{m}{2k} \big(n \omega \tilde Y^A\big)^{m-2k}\big(p_2^A\big)^{2k}\label{X4}
\end{gather}
and
\begin{gather}
X_5 = \sum_{k=0}^{[\frac{n-1}{2}]}(-1)^k\binom{n}{2k+1} \big(m\omega \tilde X^A\big)^{n-2k-1} \big(p_1^A\big)^{2k+1}\sum_{k=0}^{[\frac{m}{2}]}(-1)^k\binom{m}{2k}\big(n\omega \tilde Y^A\big)^{m-2k} \big(p_2^A\big)^{2k}\nonumber\\
\hphantom{X_5 =}{} - \sum_{k=0}^{[\frac{n}{2}]}(-1)^k\binom{n}{2k}\big(m\omega \tilde X^A\big)^{n-2k} \big(p_2^A\big)^{2k}\nonumber\\
\hphantom{X_5 =}{} \times \sum_{k=0}^{[\frac{m-1}{2}]}(-1)^k\binom{m}{2k+1} \big(n\omega \tilde Y^A\big)^{m-2k-1} \big(p_1^A\big)^{2k+1}, \label{X5}
\end{gather}
where
\begin{gather*}
\tilde X^A=x-\frac{n^2 \Omega_2 \big(p_3^A+\Omega_2x+\Omega_1 y\big)}{n^2 \Omega_2^2+ m^2\Omega_1^2},\qquad \tilde Y^A=y-\frac{m^2 \Omega_1\big(p_3^A+\Omega_2x+\Omega_1 y\big)}{n^2\Omega_2^2+ m^2\Omega_1^2}.
\end{gather*}

Since $I_1$, $I_2$, $I_3$ are independent integrals for the oscillator system, by applying the chain rule we see that the integrals $X_0$, $X_1$, $X_2$, $X_4$ form a set of independent integrals for the original system, where $X_0=p_3$. Moreover, if the gauge is chosen as in \eqref{CaseA2vecpot}, none of them depends on the~$z$ variable, while~$X_3$ does. Therefore the five integrals $X_0$, $X_1$, $X_2$, $X_3$, $X_4$ are independent. This implies the maximally superintegrability of the system~\eqref{mageff}. Notice that the same argument also applies to $X_0$, $X_1$, $X_2$, $X_3$, $X_5$.

Let us now discuss the order of the new integrals $X_4$, $X_5$. From the expressions~\eqref{X4} and~\eqref{X5}, it is clear that their order is at most $m+n$. However from~\eqref{X4} we see that the terms of order $m+n$ in $X_4$ are only of the form
\begin{gather}
 \alpha_k \gamma_j p_1^{2k}p_2^{2j} p_3^{n+m-2(k+j)},\qquad k=0,\ldots,\left[\frac{n}{2}\right],\qquad j=0,\ldots,\left[\frac{m}{2}\right]\label{higher1}
\end{gather}
or
\begin{gather}
\beta_k \delta_j p_1^{2k+1}p_2^{2j+1} p_3^{n+m-2(k+j+1)},\qquad k=0,\ldots,\left[\frac{n-1}{2}\right],\qquad j=0,\ldots,\left[\frac{m-1}{2}\right],\label{higher2}
\end{gather}
where $\alpha_j$, $\beta_j$, $\gamma_j$, $\delta_j$ are coefficients whose explicit expression can be found through~\eqref{X4}. We can eliminate all the terms of the form~\eqref{higher1} by subtracting the integrals
\begin{gather}\label{higher1int}
\alpha_k \gamma_j X_0^{n+m-2(k+j)} X_1^{k}X_2^{j},\qquad k=0,\ldots,\left[\frac{n}{2}\right],\qquad j=0,\ldots,\left[\frac{m}{2}\right].
\end{gather}
Similarly, we can subtract
\begin{gather}\label{higher2int}
\frac{\beta_k \delta_j}{2} X_0^{n+m-2(k+j+1)} X_1^{k}X_2^{j}\left( \frac{\Omega_2}{\kappa^2 \Omega_1 } \big( X_3^2-X_1\big)-\kappa^2 \frac{\Omega_1}{\Omega_2} X_2\right),
\end{gather}
where $k=0,\ldots,\big[\frac{n-1}{2}\big]$, $j=0,\ldots,\big[\frac{m-1}{2}\big]$, and eliminate all the terms \eqref{higher2}. Therefore the order of the integral $X_4$ can be reduced to $m+n-1$. By construction the terms of order $m+n-1$ of the reduced integral $\tilde{X}_4$ take the form of products of $m+n-1$ linear momenta and one coordinate. Since the highest order terms of any integral must belong to the enveloping algebra of the Euclidean algebra~\cite{MSW, MiPoWin}, we deduce that each of the highest order terms of~$\tilde{X}_4$ is a~product of $m+n-2$ linear momenta and one angular momentum. Since the leading order terms of all the other independent integrals $X_0$, $X_1$, $X_2$, $X_3$ contain only linear momenta, it is not possible to further reduce the order of the integral~$\tilde{X}_4$ by polynomial combinations of the other integrals.

Concerning the order of $X_5$, we notice that this integral contains the highest order terms of the type
\begin{gather*} %\label{xox1x21}
 \alpha_k \gamma_j p_2^{2k}p_1^{2j+1} p_3^{n-2(k+j)-1},\qquad k=0,\ldots,\left[\frac{m}{2}\right],\qquad j=0,\ldots,\left[\frac{n-1}{2}\right]
\end{gather*}
or
\begin{gather*}
 \beta_k \delta_j p_1^{2k+1}p_2^{2j} p_3^{n-2(k+j)-1},\qquad k=0,\ldots,\left[\frac{m-1}{2}\right],\qquad j=0,\ldots,\left[\frac{n}{2}\right],%\label{xox1x22}
\end{gather*}
whose order as polynomials in $p_1$ and $p_2$ is odd. Therefore, it is not possible to eliminate them by polynomial combinations of the other integrals. As far as we can see, the order of $X_5$ cannot be reduced and it is $m+n$.

\section[Example: $n=2$, $m=3$]{Example: $\boldsymbol{n=2}$, $\boldsymbol{m=3}$} \label{sec: example}

In order to illustrate the concepts and general results introduced above let us consider a particular nontrivial example. We choose the constants $n=2$ and $m=3$, i.e., $\kappa=\frac{3}{2}$. Thus the Hamiltonian of the 2 d.o.f.\ oscillator~\eqref{H2dof} reads
\begin{gather*} %\label{H2dof23}
H_2=\frac12\big(P_1^2+P_2^2\big)+ \frac12 \omega^2\big(9 X^2+4 Y^2\big), \qquad \omega^2= \frac{1}{4} \Omega_1^2+ \frac{1}{9} \Omega_2^2.
\end{gather*}
The integral $X_4$ is of order $n+m-1=4$. Thus its leading order terms are fourth order terms in the enveloping algebra of the Euclidean algebra, linear in angular momenta~$l_j$ and cubic in linear momenta~$p_j$. Explicitly, they are as follows
\begin{gather}
\nonumber X_4^{\mathrm{(h.o.)}} = \frac{1}{\sqrt{9 \Omega_1^2+4 \Omega_2^2}} \bigg( \left( \frac{16 \Omega_2^3}{9 \Omega_1}+4 \Omega_1 \Omega_2 \right) l_2 p_2^2 p_3
 - 4 \Omega_1 \Omega_2 (3 l_2 p_3+8 l_3 p_2 ) p_3^2 \\
 \hphantom{X_4^{\mathrm{(h.o.)}} =}{} -\big( 4 \Omega_2^2 +9 \Omega_1^2 \big) (l_1 p_3+l_3 p_1 ) p_2^2 +27 \Omega_1^2 (l_1 p_3+l_3 p_1 t) p_3^2\bigg).\label{X4expl}
\end{gather}
The highest order terms take the same form also when $X_4$ is expressed in a gauge covariant way, using $p_j^A= p_j+A_j(\vec x)$ and $l_j^A=\sum_{k,l} \epsilon_{jkl} x_k p_l^A$ instead of $p_j$, $l_j$.

For the sake of completeness let us also write down rather unwieldy expression for the lower order terms (for our fixed gauge~\eqref{CaseA2vecpot} since the gauge covariant expression is even more cumbersome)
\begin{gather*}
X_4-X_4^{\mathrm{(h.o.)}} = 2 \Omega_1 \tau y^2 p_1^2 p_3 -2 \tau \left(3 \Omega_1 x+\frac{8}{9} \Omega_2 y\right) y p_1 p_2 p_3-\frac{8 \Omega_2 \tau}{9} y z p_1 p_3^2 \\
\hphantom{X_4-X_4^{\mathrm{(h.o.)}} =}{}
+\tau \left( \frac{\Omega_1}{2} \big( 9 x^2+ y^2 - z^2\big) +2 \Omega_2 x y +\frac{2}{9} \frac{\Omega_2^2}{\Omega_1} \big(x^2- z^2\big) \right) p_2^2 p_3 \\
\hphantom{X_4-X_4^{\mathrm{(h.o.)}} =}{}
-\frac{1}{2 \tau} \left( 27\left( x^2-\frac{1}{3} y^2-z^2\right) \Omega_1^3-36 \Omega_1^2 \Omega_2 x y \right. \\
\left. \hphantom{X_4-X_4^{\mathrm{(h.o.)}} =}{} + 4 \Omega_2^2 \Omega_1 \big(3 x^2+4 y^2 - 3 z^2\big)-\frac{64 \Omega_2^3}{9} x y \right) p_3^3 \\
\hphantom{X_4-X_4^{\mathrm{(h.o.)}} =}{}
-2 \Omega_1 \tau y z p_2 p_3^2-\frac{\tau^3}{27} y^3 p_1^2+\frac{\tau^3}{3} x y^2 p_1 p_2+\frac{4 \Omega_2 \tau^3}{81 \Omega_1} y^2 z p_1 p_3 \\
\hphantom{X_4-X_4^{\mathrm{(h.o.)}} =}{}
- \frac{\tau^3}{4} x^2 y p_2^2+ \frac{\tau^3}{9} y^2 z p_2 p_3 \\
\hphantom{X_4-X_4^{\mathrm{(h.o.)}} =}{}
-\tau \left( \Omega_1^2 \left( 9\frac{x^2}{4}+2 y^2 - z^2\right) +\frac{4 \Omega_2^2}{9} \left(x^2-\frac{1}{3} y^2-z^2\right) +\frac{16 \Omega_2^3}{81 \Omega_1} x y \right) y p_3^2 \\
\hphantom{X_4-X_4^{\mathrm{(h.o.)}} =}{}
+ \frac{1}{18 \Omega_1} \tau^3 \left( \left( \Omega_1 y- \frac{2}{3} \Omega_2 x \right)^2
 - \left( \Omega_1^2 + \frac{4}{9} \Omega_2^2 \right) z^2\right) y^2 p_3+\frac{\tau^5}{108} y^3 x^2,
\end{gather*}
where $\tau=\sqrt{9 \Omega_1^2+4 \Omega_2^2}=6 \omega$.

\begin{figure}[t!]\centering
\includegraphics[width=60mm]{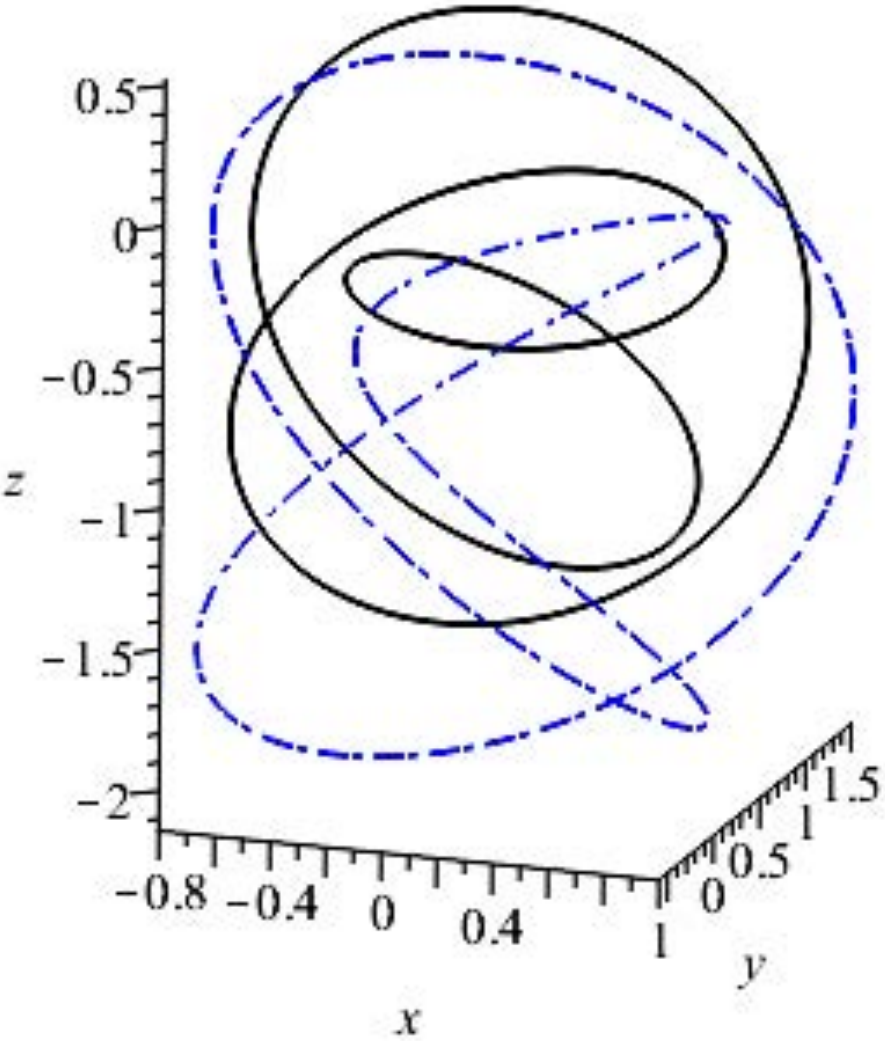}
\caption{Sample trajectories for $n=2$, $m=3$, $\vec x_0=(1,0,0)$, $\vec p_0=\big(0,1,\frac{1}{2}\big)$ and $\Omega_1 =1$, $\Omega_2 =\frac{3}{2}$ (solid line) versus $\Omega_1 =1$, $\Omega_2 =\frac{1}{2}$ (dashed line).}\label{trajectoryn2m3}
\end{figure}

Sample trajectories for two different choices of the frequencies $\Omega_{1,2}$ are shown in Fig.~\ref{trajectoryn2m3}.

\section{Conclusions} \label{sec: conclusions}

We have demonstrated in a constructive way that the classical system given by the potentials~\eqref{mageff} is maximally superintegrable whenever the parameters satisfy the rationality constraint~\eqref{Sk}. The constructed fifth independent integral is polynomial in the momenta and coordinates and it is of order~$m+n-1$ where $m$ and $n$ are incommensurable integers such that $S=\frac{ m^2 \Omega_1}{ n^2 \Omega_2}$. Its leading order terms contain angular momenta, in contrast with all the other, previously known integrals for the system~\eqref{mageff}.

The explicit form of the integral $X_4$ given as the expression~\eqref{X4} minus terms of the form~\eqref{higher1int} and~\eqref{higher2int} is unfortunately rather complicated, cf.~\eqref{X4expl}. We were not yet able to obtain any better insight into the structure of the monomials in~\eqref{X4expl}. In particular we would like to be able to predict the monomials appearing in the highest order terms for arbitrary~$m$,~$n$, together with relations between their coefficients. This understanding should be postponed to future work.

Up to this point our analysis was purely classical. Thus a natural question arises whether its results can be taken over into the quantum case. We notice that the quantum analogues of the expressions~$z_1$ and~$z_2$
\begin{gather*}
\hat{z}_1= i \hat{P}_1+m \omega \hat{X}, \qquad \hat{z}_2= i \hat{P}_2+n \omega \hat{Y}
\end{gather*}
satisfy $[\hat{z}_1,\hat{z}_2]=0$. Thus the hermitean expressions
\begin{gather}\label{2dI3I4}
\hat{I}_3= \frac{1}{2} \big( \hat{z}_1^n \big(\hat{z}_2^\dagger\big)^m+\hat{z}_2^m \big(\hat{z}_1^\dagger\big)^n \big), \qquad \hat{I}_4= \frac{1}{2 i} \big( \hat{z}_1^n \big(\hat{z}_2^\dagger\big)^m - \hat{z}_2^m \big(\hat{z}_1^\dagger\big)^n \big)
\end{gather}
are again integrals of motion,
\begin{gather*}
\big[\hat{H},\hat{I_3}\big]=0,\qquad \big[\hat{H},\hat{I_4}\big]=0
\end{gather*}
(this claim can be also verified directly through a simple commutator evaluation, see also~\cite{JauHi}). Thus the integrals of the 2 d.o.f.\ anisotropic oscillator are preserved by the quantization although their explicit expression as polynomials in $\hat{X}$, $\hat{Y}$ and $\hat{P}_{1,2}$ needs to be symmetrized due to presence of terms involving the same component of both the coordinate and the momentum, e.g.,~$\hat{X}$ and~$\hat{P_1}$ in~$\hat{z}_1$.

In order to return to the system~\eqref{mageff} we assume that the gauge is fixed as in~\eqref{CaseA2vecpot}. We notice that the Hamiltonian as well as the integrals~\eqref{X1X2} and~\eqref{X3} contain only commuting terms in each of their monomials, thus can be taken into quantum mechanics without any need for symmetrization. In the substitution
\begin{gather}\label{quantXYsubs}
\widehat{X}=\hat{x}-\frac{n^2 \Omega_2 \hat{p_3}}{n^2 \Omega_2^2+ m^2\Omega_1^2},\qquad \widehat{Y}=\hat{y}-\frac{m^2 \Omega_1 \hat{p_3}}{n^2\Omega_2^2+ m^2\Omega_1^2}
\end{gather}
we have only commuting variables $\hat{x}$, $\hat{y}$ and $\hat{p_3}$. The momentum $\hat{p_3}$ also commutes with~$\hat{p}_1$ and~$\hat{p}_2$. Thus substituting~\eqref{quantXYsubs} into the expressions~\eqref{2dI3I4} for $\hat{I}_3$ and $\hat{I}_4$ one can directly obtain the quantum integrals
\begin{gather}\label{quantX4}
\hat{X}_4=\frac{1}{2} \left( \left( i \hat{p}_1 -\frac{\Omega_2 \hat{p_3}}{m \omega}+ m \omega\hat{x}\right)^n \left( - i \hat{p}_2 -\frac{\Omega_1 \hat{p_3}}{n \omega}+ n \omega\hat{y}\right)^m+ \text{h.c.}\right)
\end{gather} and
\begin{gather}\label{quantX5}
\hat{X}_5=\frac{1}{2i} \left( \left( i \hat{p}_1 -\frac{\Omega_2 \hat{p_3}}{m \omega}+ m \omega\hat{x}\right)^n \left( - i \hat{p}_2 -\frac{\Omega_1 \hat{p_3}}{n \omega}+ n \omega\hat{y}\right)^m- \text{h.c.}\right),
\end{gather}
where ``h.c.'' stands for hermitean conjugate. Expanding the powers in~\eqref{quantX4} and~\eqref{quantX5} one obtains quantum analogues of equations~\eqref{X4} and~\eqref{X5} as their properly symmetrized versions.

 Also the argument concerning the lowering of the order of the integral $X_4$ remains the same in the quantum case, thus the integral $\hat{X}_4$ makes the quantum system maximally superintegrable of order $(m+n-1)$.

Let us notice that in accordance with~\cite{BeChaRas} and~\cite{ShaBaMe} both the Hamilton--Jacobi and the Schr\"o\-din\-ger equations separate in Cartesian coordinates. E.g., the Hamilton's principal function $S(\vec x,t)$ can be written as
\begin{gather*}
S(\vec x,t)=-E t + K_3 z+ S_1(x)+S_2(y),
\end{gather*}
where the functions $S_{1,2}$ are solutions of
\begin{gather*}
S_1'(x) = \pm \sqrt{-\big(\kappa^2 \Omega_1^2+\Omega_2^2\big) x^2+2 K_3 x \Omega_2+2 K_1},\\
S_2'(y) = \pm\sqrt{-\left(\Omega_1^2+\frac{\Omega_2^2}{\kappa^2}\right) y^2+2 K_3 y \Omega_1-K_3^2-2 K_1-2 E},
\end{gather*}
expressible in terms of square roots and inverse trigonometric functions. Whether it separates also in some other coordinate system, i.e., whether the maximally superintegrable system~\eqref{mageff} is multiseparable, remains to our knowledge an open question.

Our considerations are by construction nonrelativistic. We mention that some of the nonrelativistic superintegrable systems with magnetic fields give rise also to their superintegrable relativistic versions, as was observed in~\cite{HeiIld}, e.g., in the case of helical undulator. Due to the complicated structure of the integral~\eqref{X4} we are presently unable to construct its relativistic version and thus we do not know whether the relativistic analogue of the system~\eqref{mageff} is also superintegrable.

\subsection*{Acknowledgments}
This research was supported by the Grant Agency of the Czech Republic, project 17-11805S. The authors thank Pavel Winternitz for discussions on the subject of this paper.

\pdfbookmark[1]{References}{ref}
\LastPageEnding

\end{document}